\documentclass[aps,pra,showpacs,floatfix,nofootinbib,11pt,superscriptaddress]{revtex4-1}
\usepackage[colorlinks,linkcolor=blue,citecolor=blue,urlcolor=blue,breaklinks=true]{hyperref}
\usepackage[T1]{fontenc}
\usepackage[utf8]{inputenc}
\usepackage[]{graphicx}
\usepackage[intlimits]{amsmath}
\usepackage{float}
\usepackage{amssymb,soul,physics,enumitem,cleveref,xfrac}
\usepackage{siunitx}
\usepackage{textcomp}
\usepackage{caption}
\usepackage{subcaption}
\usepackage[table]{xcolor}
\usepackage{todonotes}

\newcommand{\kv}{{\vb{k}}}
\newcommand{\modeIndex}{{\kv,\lambda}}
\newcommand{\atom}{_\text{A}}
\newcommand{\area}{\mathcal{A}}
\newcommand{\epsNot}{\ensuremath{\varepsilon_0}}
\newcommand{\Gauss}{_\text{G}}

\begin{document}

%\title{Coupling of a single photon emitted from a Bose-Einstein condensate into a single-mode fiber}
%\title{Microwave to optical transducer via a Bose-Einstein condensate of atoms}
\title{{Collection efficiency of optical photons generated from microwave excitations of a Bose-Einstein condensate}}

\author{Árpád Kurkó}
\email{curko.arpad@wigner.hu}
\affiliation{Wigner Research Centre for Physics, H-1525 Budapest, P.O. Box 49., Hungary}

\author{Peter Domokos}
%\email{domokos.peter@wigner.hu}
\affiliation{Wigner Research Centre for Physics, H-1525 Budapest, P.O. Box 49., Hungary}

\author{David Petrosyan}
%\email{dap@iesl.forth.gr}
\affiliation{Institute of Electronic Structure and Laser, Foundation for Research and Technology – Hellas, 
GR-70013 Heraklion, Crete, Greece}

\author{András Vukics}
\email{Corresponding author: vukics.andras@wigner.hu}
\affiliation{Wigner Research Centre for Physics, H-1525 Budapest, P.O. Box 49., Hungary}

\date{\today}

\begin{abstract}
Stimulated Raman scattering on $\Lambda$-atoms is a promising tool for transducing microwave photons to optical photons. We consider an atomic Bose-Einstein condensate whose large phase-space density highly amplifies the coupling to the microwave field. Due to momentum transfer to the condensate, stimulated photon scattering can occur outside of the phase-matched direction, which can be used to separate the converted photons from the strong Raman readout pulse. Conversely, in the phase-matched direction, superradiant scattering due to bosonic enhancement leads to increased efficiency of the microwave to optical conversion. We determine the optimal conditions for the emitted optical photons to be collected into the guided modes of optical fibers.
\end{abstract}


\maketitle

\section{Introduction}

A quantum network consists of a set of quantum processing and storage nodes distributed at different locations and connected by optical fibers that transmit quantum information with photons \cite{Kimble2008,VanMeter2014,Awschalom2021}. Various platforms with experimentally verified relevant quantum capabilities have been proposed for the implementation of the nodes. These include nuclear magnetic resonance systems \cite{Vandersypen2005}, single trapped ions \cite{Schupp2021}, neutral atoms in optical lattices \cite{Heinz2020}, single atoms in optical cavities \cite{Reiserer2015}, quantum dots \cite{Lodahl2015,Senellart2017}, color centers in crystals \cite{Kalb2017}, and perhaps the most promising, superconducting circuits \cite{Arute2019,Blais2021}. The latter platform operates in the microwave regime, and quantum-coherent microwave-to-optical converters \cite{Lauk2020} will play a key role in the realization of a quantum network with superconducting-circuit nodes \cite{Forsch2020}. This is, however, a daunting task, since for the microwave to optical conversion at a single photon level, strong and coherent coupling is needed between quantum degrees of freedom differing by many orders of magnitude in energy. The most promising recent experiments employed hybrid systems to realize transducers, demonstrating remarkable achievements, including bidirectional operation \cite{Vainsencher2016}, coherent coupling \cite{Balram2016,Lambert2020}, and efficient conversion \cite{Higginbotham2018,Jiang2020}. {Moreover, there has also been recent progress in non-AMO systems realizing microwave-to-optical transducers by means of radiation pressure \cite{Mirhosseini2020,Arnold2020}. }

Neutral alkali or alkaline-earth atoms have strong optical transitions and also microwave resonances between hyperfine sublevels, and thereby provide a natural platform for realizing a quantum transducer at a single-photon level \cite{Kurko2021}. In many situations of interest, such atoms can be modeled as three-level systems with a $\Lambda$ configuration of levels. A single microwave photon \cite{Peng2016,Zhou2020} can be converted into a spin-wave excitation of the atomic hyperfine sublevels, which, in turn, can be transferred to a single optical photon in a stimulated Raman process that is inherently reversible \cite{Fleischhauer2005,Hammerer2010}. In Refs. \cite{Covey2019,Petrosyan2019}, Rydberg transitions of the atoms have been proposed as an alternative route to achieve strong dipole coupling to a microwave field. 

In the  $\Lambda$ configuration, a hyperfine transition of a single atom does not have sufficient interaction strength with a microwave photon to make a practically useful transducer. The natural mitigation is to use large ensembles of atoms \cite{Petrosyan2019,Kurko2021}. In this paper we  study the microwave to optical conversion realized by ultracold atoms that form a degenerate quantum gas – Bose-Einstein condensate (BEC) – for increased phase-space density {and large coherence length.} Trapping an atomic BEC in the vicinity of a superconducting waveguide resonator and coupling the hyperfine atomic states to the microwave resonator field \cite{Verdu2009,Henschel2010} has already been achieved \cite{Hattermann2017}.

Already in the early days of Bose-Einstein condensation experiments, two essential features of light scattering on a BEC was noted: superradiance due to bosonic enhancement, and the possibility for the light to excite density waves that extend {coherently} over the whole atomic sample \cite{Inouye1999,Stenger1999}. We note that phenomena of the same physical origin have more recently been found to lead to dynamical phase transitions and exotic phases when the BEC interacts with a resonator field \cite{Slama2007,Ritsch2013,Ostermann2016,Muniz2020,kessler2020continuous,ferri2021emerging,Marsh2021,Mivehvar2021} instead of the free-space electromagnetic field, whereas the quantum state of the BEC can be imprinted on that of the light \cite{mekhov2007probing,Mekhov2009,Keeling2010}.

Here we consider a BEC as a transducer based on the stimulated Raman process. Since the converted optical photons are intended for applications in quantum communication, we consider how efficiently such photons can be coupled into guided Gaussian modes focused into an optical fiber by a paraxial optical array \cite{Kurko2021}. {We focus on the spatial profile of the generated radiation, which is encoded not only in the spatial dependence of the readout Gaussian mode, but also in the overlap integrals between the condensate state at different stages. As the core of our work consists of the analysis of the geometrical factors, we formulated our results independently of the explicit form of the time-dependent internal dynamics, paying attention to the incoupling efficiency, and not to the transduction one. Though the internal dynamics is formally similar to the ones encountered in STIRAP or EIT, the continuum of intermediary spatial BEC states leads to an intractable problem. Our results will be thus relative to that of a single-atom scheme, referring to the consequences of having many atoms with the spatial extension of a BEC cloud.}

Since the light scattering can create excitations in the BEC that are associated with atomic momentum, the condensate participates in the momentum balance of the stimulated Raman process. In particular, the BEC can take away almost arbitrary momentum from the photonic part of the process, without significantly altering its energetics. This is because the dispersion relation of the BEC excitations is extremely flat on the momentum scale relevant to optical photons, due to the large atomic mass. The photon scattering can then occur in the directions other than the phase-matched one determined by the readout light. This side-scattering has an intensity proportional to the number of atoms in the condensate. For the forward scattering case, with the photon emitted in the phase matched direction, however, we find a superradiant behavior, with the condensate returning to the ground state, and the scattered intensity being proportional to the square of the atom number in the BEC. Exact momentum conservation can be violated because the BEC state is not an atomic momentum eigenstate, and a Gaussian beam is not a momentum eigenstate for the light. We identify situations where, due to these weak violations of the momentum conservation, the direction of the maximum of even the forward-scattered intensity deviates from the direction of the readout pulse.

This paper is organized as follows. In \cref{sec:4waveMixing}, we introduce the scheme mixing the microwave field, the optical readout pulse, the generated optical photon, and the BEC excitation. We use a second-quantized description in which the equations of motion can be written straightforwardly in the single-excitation subspace. In \cref{sec:GaussianRadiation}, we introduce our theory for emission of the converted optical photon into paraxial guided (Gaussian) modes. Since for a given direction such modes still form a broadband 1D continuum around the frequency of the emitted radiation, the radiated intensity can be written in the Born-Markov approximation, similarly to the free-space spontaneous emission. We distinguish two cases: (A) side-scattering with the BEC left with a single free-particle excitation, and (B) forward-scattering with the BEC returning to its initial state. In \cref{sec:Discussion}, we present our numerical results and discuss the findings.

\section{The four-wave mixing scheme}
\label{sec:4waveMixing}

\begin{figure}
 \includegraphics[width=.8\linewidth]{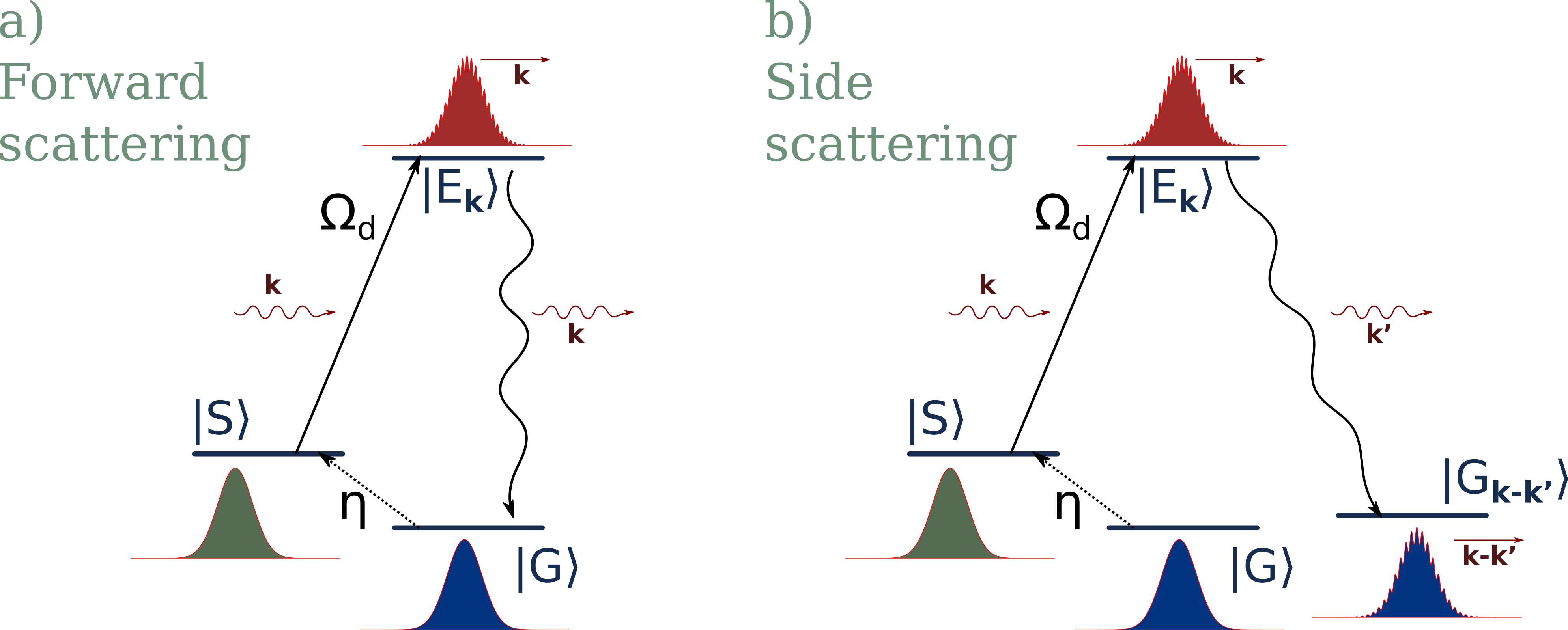}
 \caption{Level scheme depicting the many-body states available to the system starting from $\ket{G} \equiv \ket{0, 0, 0}$. 
   The other states are defined as $\ket S \equiv\ket{0,1,0}$, $\ket{E_\kv}\equiv\ket{0,0,1_\kv}$, and $\ket{G_\kv}\equiv\ket{1_{\kv},0,0}$. 
   Black lines with arrows show the transitions induced by the interaction with the microwave cavity, 
   the drive, and the free electromagnetic fields.}
 \label{fig:levelScheme}
\end{figure}

We consider a Bose-Einstein condensate of atoms with the ground state $\ket{g}$, a hyperfine state in the ground-state manifold $\ket{s}$,
and an excited state $\ket{e}$, in the $\Lambda$ configuration of levels, see \cref{fig:levelScheme}.  
Transition $\ket{g} \to \ket{s}$ is coupled to a microwave resonator mode $\hat c$ of frequency $\omega_\text{c}$ with coupling strength $\eta$. 
The hyperfine magnetic sub-states $\ket{g}$ and $\ket{s}$ are coupled by electric dipole transitions to the excited state $\ket{e}$. 
An external laser of frequency $\omega_{\mathrm{d}}$ resonantly drives the transition $\ket{s} \to \ket{e}$ with Rabi frequency $\Omega_{\mathrm{d}}$.
The transition $\ket{g} \to \ket{e}$ is coupled to the free-space modes of the electromagnetic field $ \hat{a}_\modeIndex$ of frequency $\omega_k$ 
with strength $g_{k,\lambda}$. Setting $\hbar=1$, the single-atom Hamiltonian is
\begin{multline}
\label{eq:Ham1}
H = \frac{\vb P^2}{2M}+ U(\vb r)\,\qty(\ket{g}\bra{g}+\ket{s}\bra{s}) + \omega_{gs}\ket s\bra s + \omega_{ge}\ket {e} \bra {e} \\
+ \bigg(\eta\,\hat c\,e^{-i\omega_\text{c}t}\ket s\bra g + \Omega_\text{d}\,e^{-i\omega_\text{d}t}\,e^{i\vb k_\text{d} \vb r}\ket {e}\bra s + \sum_\modeIndex g_{k, \lambda}(\vb r)\,\hat a_\modeIndex\,e^{-i\omega_k t}\ket {e}\bra {g} + \text{h.c.}\bigg) \, ,
\end{multline}
where $ \vb P^2 / 2M $ is the kinetic energy of the atom that it can acquire from the momentum transfer from the electromagnetic radiation, and $U(\vb r) $ is an optical dipole trapping potential for the atom in states $\ket{g}$ and $\ket{s}$, while the atom in the excited state is assumed free. The diagonal terms of the Hamiltonian proportional to $ \omega_{gs} $ and $ \omega_{ge} $ can be transformed out in a rotating frame, leading to 
\begin{multline}
\label{eq:Ham2}
H = \frac{\vb P^2}{2M} + U(\vb r)\,\qty(\ket{g}\bra{g}+\ket{s}\bra{s}) \\+ \bigg( \eta\,\hat c\,\ket s\bra g + \Omega_\text{d}\,e^{i\vb k_\text{d} \vb r}\ket {e}\bra s + \sum_\modeIndex g_{k, \lambda}(\vb r)\,\hat a_\modeIndex\,e^{-i(\omega_k-\omega_{ge}) t}\ket {e}\bra {g} + \text{h.c.} \bigg) \, ,
\end{multline}
where we assumed resonant interactions $ \omega_\text{c} = \omega_{gs} $ and $ \omega_\text{d} = \omega_{se} $.

%%%%%%%%%%%%%%%%%%%%%%%%%%%%%%%%%%%%%%%%%%
%%%%%%%%%%%%%%%%%%%%%%%%%%%%%%%%%%%%%%%%%%

The second-quantized Hamiltonian can be derived from \cref{eq:Ham2} as $\hat H=\int\dd{V} \hat\Psi^\dag(\vb r) H\, \hat\Psi(\vb r)$ using the field operator ansatz 
\begin{equation}
\label{eq:FieldOp}
\hat\Psi(\vb r) = \sqrt{N}\, \ket{g}\,\phi_{\text{BEC}} (\vb r) +  \sum_{\vb p \neq 0} \ket {g} \,\phi_{\vb p}(\vb r)\,\hat b_{g\,\vb p}  +  \sum_{\vb p} \ket s\,\phi_{\vb p}(\vb r)\, \hat b_{s\,\vb p} + \sum_{\vb p} \ket {e}\,\phi_{\vb p}(\vb r)\, \hat b_{e\,\vb p} \, ,
\end{equation}
where $N$ is the number of atoms, $\phi_{\text{BEC}} (\vb r)$ is the macroscopically populated BEC wave function normalized as $\int\dd{V} \phi_{\text{BEC}} (\vb r) =1$, 
$ \hat b_{g\,\vb p} $, $ \hat b_{s\,\vb p} $ and $ \hat b_{e\,\vb p} $ are the annihilation operators corresponding to states with atoms in the internal states $ \ket {g} $, $ \ket s $, $ \ket {e}$ and with the external degrees of freedom $ \vb p $, while $ \phi_{\vb p}(\vb r) \propto e^{i \vb p \cdot \vb r}$ are the corresponding motional wave functions which are the momentum eigenfunctions. The condensate wave function $ \phi_{\text{BEC}} (\vb r)$ includes the atom-atom interaction, and it is a broad function filling the trap and having large overlap with the single-particle ground state of the trap and with the zero-momentum eigenfunction $ \phi_0(\vb r) $. We can therefore take $\vb p=0$  out of the summation  over the free motional states for the ground state $ \ket{g}$. The tiny overlap with other free motional state with low momentum $\vb p \approx 0$ can safely be neglected as it has no effect on the forward photon scattering amplitude. We can also neglect the influence of the s-wave scattering of free atoms off the condensate. 
In the following, we will use in place of the zero momentum wave function $ \phi_0(\vb r)$ the BEC wave function $\phi_{\text{BEC}}(\vb r)$ also for the states $\ket{s}$ 
and $\ket{e}$. We then obtain the second-quantized Hamiltonian
\begin{multline}
\label{eq:Ham3}
\hat H = \sum_{\vb p}\frac{p^2}{2M}(\hat b_{g\,\vb p}^\dag\,\hat b_{g\,\vb p}+\hat b_{e\,\vb p}^\dag\,\hat b_{e\,\vb p}) + \left\{\sqrt{N}\,\eta\,\hat c\,\hat b_{s}^\dag + \Omega_\text{d}\,\sum_{\vb p}\hat b_{e\,\vb p}^\dag\,\hat b_{s}\,\int\dd{V}e^{i\vb k_\text{d} \vb r}\,\phi_{\vb p}^*(\vb r)\,\phi_{\text{BEC}}(\vb r)\right.
\\+ \sqrt{N}\sum_{\vb p} \sum_\modeIndex \,\hat a_\modeIndex\,e^{-i(\omega_k-\omega_{ge}) t} \hat b_{e\,\vb p}^\dag\, \int\dd{V} g_{k, \lambda}(\vb r)\,\phi_{\vb p}^*(\vb r)\,\phi_{\text{BEC}}(\vb r)
\\ \left. + \sum_{\substack{\vb p \\ \vb p'\neq 0}} \sum_\modeIndex \,\hat a_\modeIndex\,e^{-i(\omega_k-\omega_{ge}) t} \hat b_{e\,\vb p}^\dag\,\hat b_{g\,\vb p'} \int\dd{V} g_{k, \lambda}(\vb r)\,\phi_{\vb p}^*(\vb r)\,\phi_{\vb p'}(\vb r) + \text{h.c.}\right\}\; ,
\end{multline}
where we used that the motional wave functions are the orthonormal eigenfunctions of the kinetic energy, 
and neglected the momentum transfer for the microwave transition $\ket{g} \to \ket{s}$ leading to that atoms in state $\ket{s}$
have non-vanishing wavefunction $\phi_0(\vb r) \approx \phi_{\text{BEC}}(\vb r)$.
%The ground state of a BEC in a cylindrically symmetric harmonic trap can be written
%\begin{equation}
%\label{eq:groundSt}
%\phi_0(\vb r) = \frac1{(2\pi)^{\sfrac34}\,\sigma\sigma_z^{\sfrac12}}\exp(-\frac{x^2+y^2}{4\sigma^2}-\frac{z^2}{4\sigma_z^2})\,,
%\end{equation}
%where $ \sigma_x=\sigma_y=\sigma $ and $ \sigma_z $ are the oscillator lengths of the trap. If the state of motion of the condensate changes, it can gain momentum from the driving field. Let us suppose that this state of the BEC can be written as a Gaussian-modulated plane wave solution
%\begin{equation}
%\label{eq:wfFree}
%\phi_{\vb p}(\vb r) = \frac1{(2\pi)^{\sfrac34}\,\sigma\sigma_z^{\sfrac12}}\exp(-\frac{x^2+y^2}{4\sigma^2}-\frac{z^2}{4\sigma_z^2} + \frac{i}{\hbar} \vb p \vb r) \, .
%\end{equation}
{Let us confine the description to the single-excitation space where the state vector of the system can be expanded as}
\begin{multline}
\label{eq:Vec}
\ket \Psi=\delta_N\ket{0,0,0}\ket{1_\text{c}}\ket 0 + \varsigma_N\ket{0,1,0}\ket{0_\text{c}}\ket 0
+\sum_{\vb p} \epsilon_{N;\vb p}\ket{0,0,1_{\vb p}}\ket{0_\text{c}}\ket 0 \\ + \sum_{\vb p' \neq 0} \sum_\modeIndex \alpha^{\vb p'}_{N;\modeIndex} \ket{1_{\vb p'},0,0}\ket{0_\text{c}}\ket{1_\modeIndex} + \sum_\modeIndex \alpha^0_{N;\modeIndex} \ket{0,0,0}\ket{0_\text{c}}\ket{1_\modeIndex}\, 
\end{multline}
where the five terms on the rhs have the following physical meanings:
\begin{enumerate}
 \item $ \ket{0,0,0}\ket{1_\text{c}}\ket 0 $ denotes the state of the system of ultracold atoms in the pure BEC state, a single photon in the microwave resonator, and no photons in the free radiation field;
 \item $ \ket{0,1,0}\ket{0_\text{c}}\ket 0 $ denotes the state of the ensemble with a single atom in the internal state $\ket{s}$ due to the absorption of the microwave photon;
 \item $ \ket{0,0,1_{\vb p}}\ket{0_\text{c}}\ket 0 $ denotes the state of the ensemble with a single atom in state $ \ket{e}$ with momentum $ \vb p $;
 \item $ \ket{1_{\vb p'},0,0}\ket{0_\text{c}}\ket{1_\modeIndex} $ denotes the state of the ensemble with an atom in state $ \ket g $ having the momentum $ \vb p' $, plus an optical photon in the mode with the wave vector $\kv$ and the polarization $\lambda$;
 \item $\ket{0,0,0}\ket{0_\text{c}}\ket{1_\modeIndex}$ denotes a state with all the atoms in the pure BEC state with no excitations, plus an optical photon in the mode with the wave vector $\kv$ and the polarization $\lambda$.
\end{enumerate}
Note that the last two states account for the two possible photon emission channels corresponding to the system of ultracold atoms returning to the BEC state $ \phi_\text{BEC}(\vb r)$ with no excitation (“forward scattering”) or to a single motional excitation state $ \phi_{\vb p' \neq 0}(\vb r) $ (“side scattering”), which is illustrated in \cref{fig:levelScheme}.

Since the kinetic energy $ \frac{p^2}{2M} $ associated with the atomic motion is negligible in comparison to the optical radiation frequencies $ \omega_\text{c} $ and $ \omega_\text{d} $,\footnote{$ \frac{p^2}{2 M} $ can be expressed as $  \frac{\omega_\text{d}^2}{2Mc^2} $, so that the ratio $\frac{p^2/2M}{\omega_\text{d}}$ is very small because of the large rest mass of the atoms.} the diagonal terms of \cref{eq:Ham3} can be taken out. Then the time evolution of the amplitudes is obtained from the Schrödinger equation $i\derivative{\ket\Psi}{t}=\hat H \ket\Psi$, with the initial conditions $ \delta_N(t=0) = 1 $ and $ \varsigma_N(0) = \epsilon_{N;\vb p}(0) = \alpha^{\vb p' \neq 0}_{N;\modeIndex}(0) = \alpha^0_{N;\modeIndex}(0) = 0 $ for all $ \vb p $, $ \vb p ' $, $ \kv $, and $ \lambda $, as 
\begin{subequations}
\label{eq:TimeEv}
\begin{align}
i \partial_t \delta_N =& \sqrt{N} \, \eta^* \varsigma_N \label{eq:TimeEv0} \\
i \partial_t \varsigma_N =& \sqrt{N} \, \eta \, \delta_N + \sum_{\vb p} \epsilon_{N;\vb p} \, \Omega_\text{d}^*\,\int\dd{V} e^{-i\vb k_\text{d} \vb r}\phi_{\vb p}(\vb r)\,\phi_\text{BEC}^*(\vb r) \label{eq:TimeEv1} \\
i \partial_t \epsilon_{N;\vb p} =& \varsigma_N \, \Omega_\text{d}\, \int\dd{V} e^{i\vb k_\text{d} \vb r}\phi_{\vb p}^*(\vb r)\,\phi_\text{BEC}(\vb r) \nonumber \\
&+ \sum_{\vb p' \neq 0} \sum_\modeIndex \,\alpha^{\vb p'}_{N;\modeIndex}\,e^{-i(\omega_k-\omega_{ge}) t} \int\dd{V} g_{k, \lambda}(\vb r) \phi_{\vb p}^*(\vb r)\, \phi_{\vb p'}(\vb r) \nonumber \\
&+ \sqrt{N} \, \sum_\modeIndex \,\alpha^{0}_{N;\modeIndex}\,e^{-i(\omega_k-\omega_{ge}) t} \int\dd{V} g_{k, \lambda}(\vb r) \phi_{\vb p}^*(\vb r)\, \phi_\text{BEC}(\vb r) \label{eq:TimeEv2}\\
i \alpha^{\vb p'\neq0}_{N;\modeIndex} &= \sum_{\vb p} \, \int_0^t\dd{t'}\epsilon_{N;\vb p}(t')\,e^{i(\omega_k-\omega_{ge}) t'} \int\dd{V} g_{k, \lambda}^*(\vb r) \phi_{\vb p}(\vb r)\,\phi_{\vb p'}^*(\vb r)\nonumber \\
i \alpha^{0}_{N;\modeIndex} &= \sqrt{N} \, \sum_{\vb p} \, \int_0^t\dd{t'}\epsilon_{N;\vb p}(t')\,e^{i(\omega_k-\omega_{ge}) t'} \int\dd{V} g_{k, \lambda}^*(\vb r) \phi_{\vb p}(\vb r)\,\phi_\text{BEC}^*(\vb r)\, . \label{eq:TimeEv3}
\end{align}
\end{subequations}
The first three equations account for the atomic dynamics, while the last two equations describe the generation of the optical radiation field from the atomic source and are integrated to obtain directly the field amplitudes. As seen from the second part of \cref{eq:TimeEv3}, apart from the nontrivial dependence on $ N $ through the amplitude $ \epsilon_{N;\vb p} $, there is a multiplicative factor of $ \sqrt{N} $ due to the bosonic enhancement of photoemission when the condensate returns to the state $ \phi_\text{BEC}(\vb r) $.

Our main focus is the spatial profile of the emitted radiation that is collected by Gaussian optics. 
In the eq. (\ref{eq:TimeEv3}) for the field amplitudes, their spatial profiles are encoded in the overlap integrals, {and in the amplitudes $ \epsilon_{N;\vb p}(t) $ which latter also depends on time.} {At this point, considering eqs. (\ref{eq:TimeEv0}-\ref{eq:TimeEv3}), there are two ways to treat this problem: the one which can handle the general case, where the time and spatial dependence of the amplitudes are not necessarily separable, and it is extremely demanding to solve it numerically, as we have a continuum set of momentum states; the other one which solves analytically the equation of motions in certain limits in such a way that the amplitudes become separable in their time and spatial variables. In this work we are going to choose the second implementation, as we are interested in how the geometrical parameters affect the incoupled photon rate, regardless of the internal dynamics of the system.} 

To this end, we separate the temporal and spatial dependence in these formulas, we have to separate the temporal and wave-vector dependence of the amplitudes $ \epsilon_{N;\vb p}(t) $. {This is a good approximation in the perturbative limit, considering a short amount of time at the beginning of the process, when the time evolution eqs. (\ref{eq:TimeEv0}-\ref{eq:TimeEv2})  can be approximated by the content of the initially fully occupied amplitude $ \delta_N $ being transferred into the initially empty amplitudes $ \varsigma_N $ and $ \epsilon_{N;\vb p} $:
\begin{subequations}
\begin{align}
i \partial_t \varsigma_N \approx& \sqrt{N} \eta(t) \, ,\label{eq:TimeEv1Pert} \\
i \partial_t \epsilon_{N;\vb p} \approx& \varsigma_N \, \Omega_\text{d}(t)\, \int\dd{V} e^{i\vb k_\text{d} \vb r}\phi_{\vb p}^*(\vb r)\,\phi_\text{BEC}(\vb r) \, , \label{eq:TimeEv2Pert}
\end{align}
\end{subequations}
where the first equation of (\ref{eq:TimeEv0}-\ref{eq:TimeEv2}) is eliminated, $ \delta_N $ is treated simply as a decaying amplitude, and we introduced the time-dependent drive $ \eta(t)$ by the product of the microwave coupling strength $ \eta $ and the depleting initial state amplitude $ \delta_N $.
We assumed that during this short period, in the time evolution of $ \varsigma_N $ and $ \epsilon_{N;\vb p} $ only the first, leading order terms will contribute. Performing the time integration, the amplitude corresponding to the intermediate state $ \lbrace \ket{e}; \vb p \rbrace $ reads
\begin{multline}
\epsilon_{N;\vb p} (t) \approx -\sqrt{N} \int_0^{t} \int_{0}^{t'} \dd{t'} \dd{t''} \Omega_\text{d}(t') \eta(t'')  \, \int\dd{V} e^{i\vb k_\text{d} \vb r}\phi_{\vb p}^*(\vb r)\,\phi_\text{BEC}(\vb r) \\
\equiv \sqrt{N} \, \epsilon(t) \, \int\dd{V} e^{i\vb k_\text{d} \vb r}\phi_{\vb p}^*(\vb r)\,\phi_\text{BEC}(\vb r) = \sqrt{N} \, \frac{\epsilon (t)}{\sqrt{V}}\int\dd{V} e^{i\vb k_{\text d }\vb r} e^{-i\vb p \vb r} \phi_\text{BEC}(\vb r)
=\sqrt{N} \, \frac{ \epsilon (t)}{\sqrt{V}} \, \tilde{\phi}_\text{BEC}(\vb p - \vb k_\text{d})\, \label{eq:Eamp}, 
\end{multline}
where we used the momentum eigenstates $ 1/\sqrt{V} \, e^{i\vb p \vb r} $ for the wavefunction of the intermediate states, we denoted the Fourier transform of $\phi_\text{BEC}$ by $\tilde{\phi}_\text{BEC}$, and we have factorized the $N$ and wave number $\vb p$ dependence out of the amplitude $ \epsilon_{N;\vb p} (t) $ by introducing $ \epsilon(t) $, depending only on time.} 

{Hereinafter, we are going to focus on the spatial profile of the generated radiation and on the atom number dependence of the transducer process: calculating the photon rate of the two channels defined in \cref{eq:TimeEv3} and coupled into an optical fiber, using the amplitude $ \epsilon_{N;\vb p} $ from \cref{eq:Eamp} for the $ \ket{g} \rightarrow  \lbrace \ket{e}; \vb p \rbrace $ part of the photon conversion.}

\section{Radiation into Gaussian modes}
\label{sec:GaussianRadiation}

\begin{figure}
\centering
\includegraphics[width=0.8\linewidth]{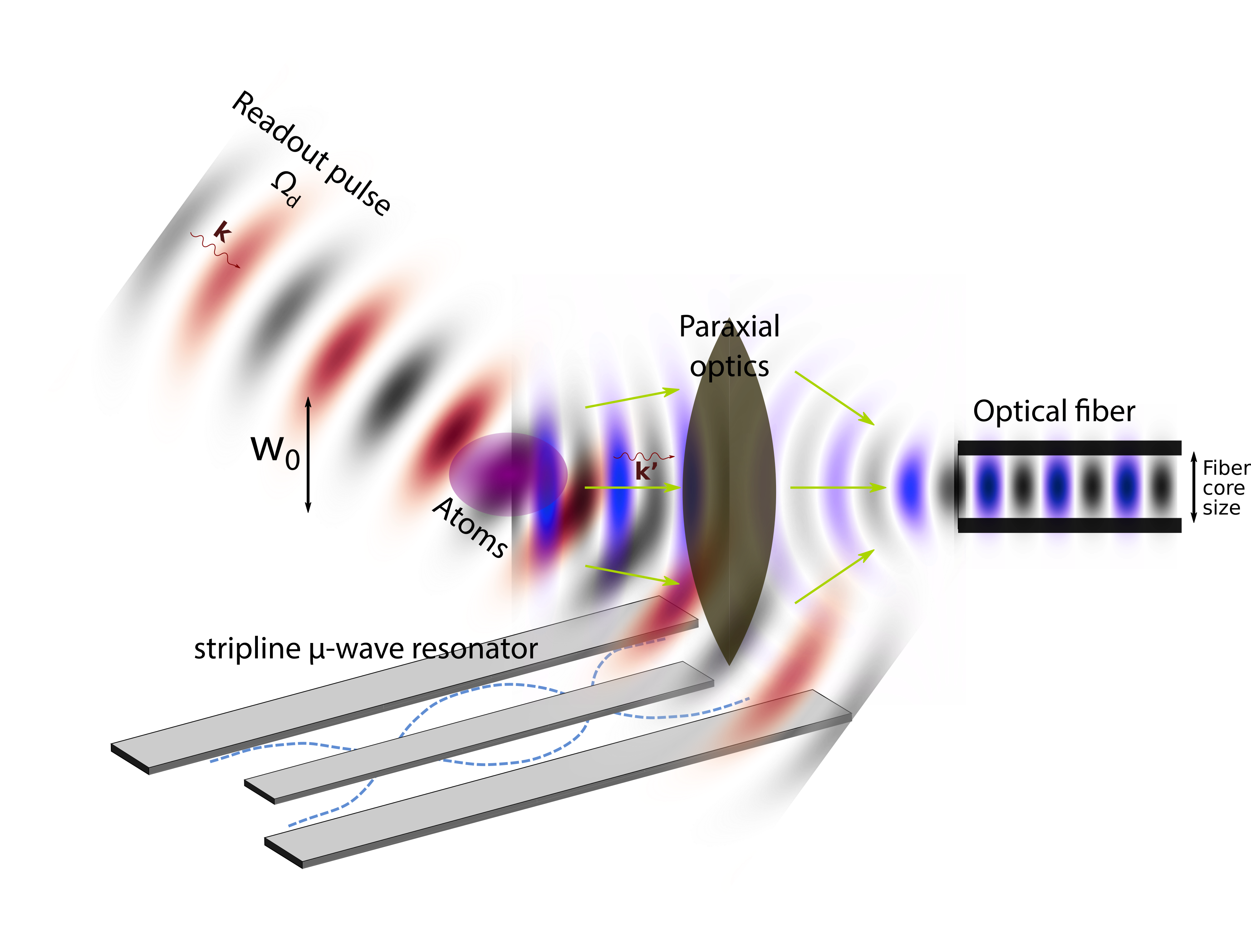}
\caption{The geometry of the system of atoms, paraxial optics and the fibre.}
\label{fig:inCoupling}
\end{figure}

The aim is to collect the emitted radiation into an optical fiber. The modes of a single-mode fiber form a one-dimensional manifold, which can be parametrized by a scalar wavenumber $k$. The important parameters are the size of the fiber core and the orientation of the fiber with respect to the axis of the driving field that determines the phase-matched emission direction. To model the incoupling process, we assume a generic paraxial optical array, which transforms a set of free, nearly paraxial beams to the strongly confined guided mode, see \cref{fig:inCoupling}. The fiber modes confined in the core of size $d_\text{core}$ are transformed in free space to Gaussian modes with waist $w_0$ determined by the parameters of the optical array at the position of the ensemble:\footnote{According to the ABCD rule of paraxial optics \cite{SalehTeich}, this transformation depends on the wavenumber, but in our case the relevant wavenumbers are very strongly confined to the vicinity of the two-photon resonance, hence we neglect this dependence.}
\begin{equation*}
d_\text{core}, k\qqtext{@ fiber entrance}\Longleftrightarrow\quad w_0, k\quad\text{@ position of the BEC.}
\end{equation*}

%%%%%%%%%%%%%%%%%%%%%%%%%%%%
%%%%%%%%%%%%%%%%%%%%%%%%%%%%

On the basis of the foregoing considerations, from the full set of free-space modes we can separate a set of guided modes with wave vectors aligned with the axis of the fiber that are coupled by the optical array into the fiber. These modes can be well approximated by Gaussian modes with wavenumber $k$ and waist $w_0$, with mode function
\begin{multline}
\label{eq:GaussianMode}
f_k (\mathbf{r}) = \frac{z_\text{R}}{q^*(z)} \exp(ik  \qty[ z + \frac{\rho^2}{2\,q^*(z)} ] ) \\
= i\frac{w_0}{w(z)}\,\exp[ikz - \frac{\rho^2}{w(z)^2} + i\frac{k\rho^2}{2R(z)} - i\phi_\text{Gouy}] \equiv \varphi_{\text G}(\vb r)\,e^{ikz}\,,
\end{multline}
where $z$ is the position along the direction of the fiber axis, $\rho=\sqrt{x^2+y^2}$ is the radial coordinate, 
$q(z) = z + i z_\text{R} $ is the complex beam parameter with the Rayleigh range $z_\text{R} = k w_0^2/2$, 
$ w(z)=w_0\sqrt{1+z^2/z_\text{R}^2} $ is the transverse beam size, $ R(z) = z + z_\text{R}^2/z$ is the radius of curvature of the wave front, 
and $ \phi_\text{Gouy} = \arctan(z/z_\text{R}) $ is the Gouy phase. 
{For a given optical array, with a well-defined beam waist $ w_0 $ and direction of the axis, we can claim that the set of Gaussian modes parametrized by the $k$ forms an orthonormal set, this being important in order to ensure the correct commutation relation for the electric field. The amplitudes of the guided modes corresponding to the $N$ and $\vb p'$ dependent free-space amplitudes $\alpha^{\vb p'\neq0}_{N;\modeIndex}$ and $ \alpha^{0}_{N;\modeIndex} $ are denoted by $\beta^{\vb p'\neq0}_{N;k,\lambda}$ and $ \beta^{0}_{N;k,\lambda} $. The $\beta$ amplitudes simply replace the $\alpha$s in \cref{eq:TimeEv} in the case when all electromagnetic processes are confined to the guided modes.}

The expectation value of the positive frequency part of the electric field that is coupled into the fiber reads 
{ 
\begin{equation}
\vb E\Gauss ^+(\mathbf{r}) = i \sum_{\vb p'} \sum_{k, \lambda}  \sqrt{\frac{\omega}{2 \epsNot V}}\, \beta^{\vb p'}_{N;k,\lambda}(t)\, \vb e_\lambda\, f_k(\mathbf{r}) \,,
\end{equation}}
where $V=\area L$ is the mode volume with $\area = \pi w_0^2/4$ being its cross-sectional area and $L$ the fictitious quantization length along the propagation axis, {and there is a sum for all subchannels labeled by all the possible momenta $ \vb p' $, the BEC acting as a momentum reservoir there is no constraint for the absorbable momentum $ \vb p' $.} The fibre-coupled modes form a broadband continuum and the field $\vb E\Gauss$ can be treated similarly to that of the three-dimensional electromagnetic vacuum surrounding the atom \cite{Domokos2002}. Irreversible photon emission from an excited-state atom into the fibre takes place, and the photon emission rate is determined by the local coupling between the atom and the modes,
\begin{equation}
\label{eq:CouplingGaussian}
g_{k, \lambda}({\vb r}\atom) =  \sqrt{ \frac{\omega}{2 \epsNot V}}\, \vb d_{ge} \vb e_{\lambda}\,  f_k ({\vb r}\atom) = g_{k, \lambda} \, f_k ({\vb r}\atom) \, . 
\end{equation}
{When the newly introduced amplitudes $\beta^{\vb p'\neq0}_{N;k,\lambda}$ and $ \beta^{0}_{N;k,\lambda} $ replace the corresponding $\alpha$s in \cref{eq:TimeEv3}, the coupling constants have to be written as \cref{eq:CouplingGaussian}.}

{The total photon numbers generated during a small time interval $ \Delta t $, denoted $ \Delta n (t) $ and $ \Delta n^0(t) $ in the side- and forward-scattering channels, respectively, read
\begin{subequations}
\begin{align}
\Delta n (t) =& \sum_{\vb p' \neq 0} \sum_{k, \lambda} | \Delta\beta^{\vb p'}_{N;k,\lambda}(t) |^2 = \sum_{\vb p' \neq 0} \sum_{k, \lambda} \left| \, \int_{t-\Delta t}^{t} \dd{t'} \beta^{\vb p'}_{N;k,\lambda}(t) \right|^2 \, \label{eq:NumPhFree}\\
\Delta n^0 (t) =& \sum_{k, \lambda} | \Delta\beta^{0}_{N;k,\lambda}(t) |^2 = \sum_{k, \lambda} \left| \, \int_{t-\Delta t}^{t} \dd{t'} \beta^{0}_{N;k,\lambda}(t) \right|^2 \, . \label{eq:NumPhGround}
\end{align}
\end{subequations}
}
%Let's consider a short time $ \Delta t $ which is long compared to the reservoir correlation time $ \tau $, but much shorter than the dynamical time scales of the atom-radiation coupling. 
%\begin{equation}
%\label{phNum}
%\Delta n(t) = \sum_{k, \lambda} \abs{\beta_\modeIndex(t)}^2\, ,
%\end{equation}
%where the time evolution of the amplitudes $\beta_\modeIndex(t)$ is  given in \cref{eq:TimeEv3}.

\subsection{Free final state of the atom}
\label{sec:ExcitedBEC}

Let us first consider the case when the ensemble of ultracold atoms excited into the state $\ket{e}$, on emitting a photon, does not return to the BEC state $ \phi_\text{BEC}(\vb r) $ but ends up in an arbitrary free state with momentum ${\vb p'}$ (side scattering). The total photon number $ \Delta n(t) $ according to \cref{eq:NumPhFree} reads 
\begin{multline}
\label{eq:na}
\Delta n(t) = \sum_{\vb p'} \sum_{k, \lambda} | \Delta\beta^{\vb p'}_{N;k,\lambda}(t) |^2 \\
= \sum_{\vb p'} \sum_{\vb p_1,\vb p_2}\,\sum_{k, \lambda} \, |g_{k, \lambda}|^2 \, 
\int_{t-\Delta t}^t\dd{t_1}\epsilon_{N; \vb p_1}^*(t_1)\,e^{i(\omega_k-\omega_{ge}) (t - t_1)} \, 
\int_{t-\Delta t}^t\dd{t_2}\epsilon_{N; \vb p_2}(t_2)\,e^{i(\omega_k-\omega_{ge}) (t_2-t)}\, \times \\
\times \int\dd{V}f_k(\vb r)\phi_{\vb p_1}^*(\vb r)\phi_{\vb p'}(\vb r) \,
\int\dd{V'}f_k^*(\vb r')\phi_{\vb p_2}(\vb r')\phi^*_{\vb p'}(\vb r') \, ,
\end{multline}
{where for the amplitudes $ \beta^{\vb p'}_{N;k,\lambda}(t) $ we used the first line of \cref{eq:TimeEv3} with the coupling constants $ g_{k, \lambda}(\vb r) $ defined in \cref{eq:CouplingGaussian}.  The coupling constant $ g_{k, \lambda} $ is a flat function of the wavenumber $ k $, whereas the overlap integrals (the last two terms of \cref{eq:na}) result in a function which has a bandwidth $ 1/\Lambda $, where $ \Lambda $ is the characteristic length corresponding to the BEC. However, the summation over all possible final momenta $\vb p'$ ensures that the fibre-coupled modes form a broadband continuum around the transition frequency $\omega_{ge}$ with a bandwidth much larger than the inverse of the short time interval $\Delta t$. We also assume weak coupling and neglect the reabsorption of photons within the cloud. The summation over the wavenumber $k$ together with the time integrals create thus a situation analogous to the calculation of spontaneous emission in the three-dimensional free-space modes. Hence we can adopt the Born-Markov approximation to evaluate the sum in Eq.~(\ref{eq:na}).  Because of the broadband summation over $k$ of the exponential terms $e^{i(\omega_k-\omega_{ge}) (t_2 - t_1)} $, only the time instant $t_1\approx t_2$ contribute to the combined summation over $k$ and the time integrals over $t_{1,2}$. The atomic amplitudes $\epsilon(t_{1,2})$ are thus taken in the same time and, furthermore, because the variation in the period $\Delta t$ is infinitesimal, both amplitudes can be replaced by $\epsilon(t)$ and taken out of the time integrals. The remaining time integrals can be carried out, one of them giving a Dirac-delta in frequency space, and the other one simply $\Delta t$, as follows:}
{\begin{multline}
\label{eq:nb}
\Delta n(t) \approx N \sum_{\vb p'}\sum_{\vb p_1, \vb p_2} \, \epsilon^*_{N; \vb p_1}(t) \epsilon_{N; \vb p_2}(t) \, \sum_{k, \lambda} |g_{k, \lambda}|^2  \int_{t-\Delta t}^t\dd{t_1}  \int_{t-\Delta t}^t\dd{t_2} e^{i(\omega_k-\omega_{ge}) (t_2-t_1)} \times \\
\times \int\dd{V}f_k(\vb r)\phi_{\vb p_1}^*(\vb r)\phi_{\vb p'}(\vb r) \,
\int\dd{V'}f_k^*(\vb r')\phi_{\vb p_2}(\vb r')\phi^*_{\vb p'}(\vb r') \\
= N \sum_{\vb p'}\sum_{\vb p_1, \vb p_2} \, \epsilon^*_{N; \vb p_1}(t) \epsilon_{N; \vb p_2}(t) \, \sum_{\lambda} \frac L{2\pi} \int\dd k\,\frac{\omega_k}{2\epsNot\,V} (\vb d_{ge} \vb e_{\lambda})^2 \Delta t 2 \pi \delta(\omega_k - \omega_{ge}) \times \\
\times \int\dd{V}f_k(\vb r)\phi_{\vb p_1}^*(\vb r)\phi_{\vb p'}(\vb r) \,
\int\dd{V'}f_k^*(\vb r')\phi_{\vb p_2}(\vb r')\phi^*_{\vb p'}(\vb r') \\
= N\,\Gamma\,\frac{\sigma\atom}{\area} \, \Delta t \, \sum_{\vb p'}\abs{\sum_{\vb p} \epsilon_{N; \vb p}(t) \int\dd{V}f_{k_{ge}}^*(\vb r)\phi_{\vb p}(\vb r)\phi^*_{\vb p'}(\vb r)}^2 \, ,
\end{multline}}
Here, the sum over the polarization $\lambda$ was simplified by assuming the optimum configuration: the driving laser is linearly polarized perpendicular to the plane of \cref{fig:inCoupling}, i.e., the plane spanned by the propagation of the driving laser beam $\vb k_\text{d}$ and the fibre-coupled modes. The induced atomic dipole moment $\vb d_{ge}$ has then the same out-of-plane direction. One of the two possible polarizations of the fibre-coupled modes is also  perpendicular to this plane, hence parallel to the dipole moment $\vb d_{ge}$. The other one is in-plane, being orthogonal and thus decoupled from the atomic dipole. The remaining derivation follows the standard Born-Markov (Wigner-Weisskopf) procedure. We identify the free-space atomic spontaneous emission rate $\Gamma = {\omega_{ge}^3\,d_{ge}^2}/({6 \pi \epsNot  c^3}) $ and the radiative cross section of the atomic dipole $\sigma\atom = 3 \lambda^2/(2\pi)$ in order to put the final result in a compact form.

Using the approximation (\ref{eq:Eamp}) for the amplitudes $ \epsilon_{N; \vb p}(t) $, the rate of photons $I=\Delta n/ \Delta t$ coupled into the fibre is 
\begin{multline}
\label{eq:rate}
I(t) =  N\,\Gamma\,\abs{\epsilon(t)}^2 \, \frac{\sigma\atom}{\area} \, \sum_{\vb p'}\abs{\frac{1}{V\sqrt{V}} \sum_{\vb p} \tilde{\phi}_\text{BEC}(\vb p - \vb k_\text{d})\int\dd{V} \varphi_\text{G}^*(\vb r) e^{i(\vb p - \vb k_{ge} - \vb p') \vb r}}^2 \\
= N\,\Gamma\,\abs{\epsilon(t)}^2 \, \frac{\sigma\atom}{\area} \, \sum_{\vb p'}\abs{\int \dd{V} \frac{1}{\sqrt{V}}\phi_\text{BEC}(\vb r)\varphi_{\text G}^*(\vb r)e^{i(\vb k_{\text d}-\vb k_{ge} - \vb p')\vb r}}^2  \\
= N \, \Gamma\,\abs{\epsilon(t)}^2 \, \frac{\sigma\atom}{\area} \, \int \dd{V} \abs{\phi_\text{BEC}(\vb r) \varphi_\text{G}(\vb r)}^2 \equiv N \, \Gamma\,\abs{\epsilon(t)}^2 \, \frac{\sigma\atom}{\area} \, \xi \,.
\end{multline}
We thus find that the photon scattering rate into the fibre is proportional to (i) the number of atoms $N$, (ii) the single-atom photon scattering rate $\Gamma\,\abs{\epsilon(t)}^2$ in the full solid angle, (iii) the ratio of the atomic radiative cross section and the cross section of the fibre coupled beam, and (iv) a geometrical factor $\xi$ which includes the non-trivial condition on the spatial matching of light beams and the atom cloud.

It is remarkable that, in the plane of the driving laser and the fibre-coupled mode, the intensity $I(t)$ is independent of the angle between the wave vector $ {\vb k_d} $ and the fibre axis. The reason is that momentum conservation in the photon scattering process is not restrictive in the case of a BEC initial state of the atomic ensemble. Due to its extremely flat dispersion relation compared to that of the photons, the BEC acts as a momentum reservoir that can absorb away any momentum mismatch without significantly altering the energetics of the photon scattering process. A broad angular distribution has a significant consequence in practical applications: the generated photon can be easily separated  from the driving laser beam, without the need for narrowband spectral filters, which necessarily degrade the coupling efficiency into the fibre. We note that this isotropic distribution (or, dipole pattern in the case of a fixed direction of the atomic dipole moments) holds only for the first photon; already the second photon, encountering the Bragg grating created by the excitation left by the first photon, will preferentially scatter into the same direction \cite{Inouye2000}.

\subsection{BEC final state of the atom}
A case of special interest is when the final state of the atoms after the photon emission coincides with the BEC state as in the beginning. Although the final state is now a single motional state $\phi_\text{BEC}$, and not a broad continuum, the scattered intensity is sizable because of the bosonic enhancement. The probability of this transition is strongly enhanced by the macroscopic part of the atomic population in the BEC state. {The generated field amplitude in the fibre is given by the second line of \cref{eq:TimeEv3}. When going through the same steps as in eqs. (\ref{eq:nb}), (\ref{eq:rate}) to get $ \Delta n^0(t)$, extra care must be taken in applying the Born-Markov approximation. Unlike the previous case of free final state, now there is no summation over the broadband continuum of momentum states $p'$, i.e., 
\begin{multline}
\label{eq:n0a}
\Delta n^0(t) = \sum_{k, \lambda} | \Delta\beta^{0}_{N;k,\lambda}(t) |^2 \\
= \sum_{\vb p_1,\vb p_2}\,\sum_{k, \lambda} \, |g_{k, \lambda}|^2 \, 
\int_{t-\Delta t}^t\dd{t_1}\epsilon_{N; \vb p_1}^*(t_1)\,e^{i(\omega_k-\omega_{ge}) (t - t_1)} \, 
\int_{t-\Delta t}^t\dd{t_2}\epsilon_{N; \vb p_2}(t_2)\,e^{i(\omega_k-\omega_{ge}) (t_2-t)}\, \times \\
\times \int\dd{V}f_k(\vb r)\phi_{\vb p_1}^*(\vb r)\phi_{\text{BEC}}(\vb r) \,
\int\dd{V'}f_k^*(\vb r')\phi_{\vb p_2}(\vb r')\phi^*_{\text{BEC}}(\vb r') \, .
\end{multline}
The wavenumber summation has a finite support which originates from the overlap integrals. Physically, this bandwidth is determined by (i) the width $ 1/ \Lambda $ of the BEC wavefunction in momentum space, and (ii) the width of the intermediate states $p_{1,2}$. This latter incorporates the drive laser pulse width and the BEC bandwidth. The Born-Markov approximation can be adopted to simplify the above equation if the characteristic time scale to generate a photon in this four-wave mixing process is much longer than the correlation time $ \tau_{c} $ where $(c \tau_c)^{-1}$ is the effective bandwidth of the fibre mode continuum just discussed. Under this condition, the intensity in this photon scattering channel reads}
\begin{multline}
\label{eq:rate0}
I_0(t) = \sum_{k, \lambda} | \Delta\beta^{0}_{N;k,\lambda}(t) |^2/\Delta t = N^2 \, \Gamma\,\abs{\epsilon(t)}^2 \, \frac{\sigma\atom}{\area} \, \abs{\int \dd{V} \abs{\phi_\text{BEC}(\vb r)}^2\varphi_{\text G}^*(\vb r)e^{i(\vb k_{\text d}-\vb k_{ge})\vb r}}^2 \\
\equiv  N^2 \, \Gamma\,\abs{\epsilon(t)}^2 \, \frac{\sigma\atom}{\area} \, \abs{\xi_0}^2 \, .
\end{multline}
Note that the intensity scales now quadratically with the number of atoms, while the geometric factor $ \xi_0 $ involves the BEC wavefunction in a different power, as compared to Eq.~(\ref{eq:rate}). Moreover, the geometric factor is no longer isotropic: there is no momentum transfer to the final BEC state that would compensate the momentum mismatch of the readout and emitted photon, so the outgoing photon has to take the momentum of the absorbed photon from the driving laser, resulting in predominantly forward scattering. {This result for the photon scattering rate allows for justifying the validity of the Born-Markov approximation {\it a posteriori}. If the photon generation process in the forward direction becomes very efficient, i.e., $I_0\tau_c \sim 1$, the Born-Markov approximation breaks down. This can happen, for example, for large number of atoms. The underlying physical picture is that the photon re-absorption cannot be neglected for the case of strong-coupling between the BEC and the optical modes. However, this is a favourable case in which the optical photon can be generated in a short timescale by means of an appropriate laser pulse. }

\subsection{Analytical approximations for the scattered intensity}
In order to proceed with analytic expressions, we approximate the BEC wavefunction by a Gaussian corresponding to the ground state of a non-interacting gas in a harmonic trap with cylindrical symmetry,
\begin{equation}
\label{eq:groundSt}
\phi_\text{BEC}(\vb r) = \frac1{(2\pi)^{\sfrac34}\,\sigma\sigma_z^{\sfrac12}}\exp(-\frac{x^2+y^2}{4\sigma^2}-\frac{z^2}{4\sigma_z^2})\,,
\end{equation}
where $ \sigma_x=\sigma_y=\sigma $ and $ \sigma_z $ are the oscillator lengths of the trap. We consider the configuration in which the axis of the optical fibre coincides with the long axis of the condensate (cf. \cref{fig:inCoupling}), characterized by the oscillator length $ \sigma_z $. In the case of a free-atom final state, the geometric factor $\xi$ reads
\begin{multline}
\label{eq:Xi}
\xi = \int\dd{V} \abs{\phi_\text{BEC}(\vb r)\varphi_\text{G}(\vb r)}^2 \\
= \frac{w_0^2}{(2\pi)^{3/2}\sigma^2 \sigma_z} \int\dd{V} \frac{1}{w(z)^2} \exp[-\frac{x^2+y^2}{2}\left( \frac{1}{\sigma^2} + \frac{1}{w(z)^2}\right) -\frac{z^2}{2\sigma_z^2} ]  \\
= \frac{w_0^2}{\sqrt{2\pi}\sigma_z}\int \dd{z} \frac{\exp(-\frac{z^2}{2\sigma_z^2})}{\sigma^2+w(z)^2} \approx  \frac{w_0^2}{\sigma^2 + w_0^2}\, ,
\end{multline}
where the last approximation is valid in the typical case when the longitudinal length $ \sigma_z $ of the condensate is much smaller than the Rayleigh range, so that the transverse beam size $ w(z) $ can be approximated by the beam waist $ w_0 $. The result shows that the geometrical factor 
$\xi \lesssim 1$. The larger the beam waist, the better the geometrical coupling factor. This is reasonable as the optics that couples into the fibre collects light from larger solid angle.

If the ensemble of ultracold atoms returns to the BEC state, the geometric factor, within the same approximations, is 
\begin{multline}
\label{eq:XiGround}
\xi_0(\theta) = \int\dd{V} \abs{\phi_\text{BEC}(\vb r)}^2\varphi_\text{G}^*(\vb r) e^{i(\vb k_\text{d}-\vb k_{ge})\vb r}\\
= \frac{z_\text{R}}{(2\pi)^{3/2}\sigma^2 \sigma_z} \int\dd{V} \frac{1}{q(z)} \exp[-\frac{x^2+y^2}{2}\left( \frac{1}{\sigma^2} + i\frac{k_{\text d}}{q(z)}\right) -\frac{z^2}{2\sigma_z^2} + ik_{\text d}(\cos\theta-1)\,z + ik_{\text d}\sin\theta\,x]\\
= \frac{z_\text{R}}{\sqrt{2\pi} \sigma_z} \int\dd{z} \frac{1}{q(z)+ik_{\text d}\sigma^2} \exp[-\frac{ik_{\text d}^2 \sigma^2q(z)}{2(q(z)+ik_{\text d}\sigma^2)}\sin^2\theta + ik_{\text d}(\cos\theta-1)\,z - \frac{z^2}{2\sigma_z^2}] \, ,	
\end{multline}
where $ \theta $ is the angle between the direction of the incoming driving field $ \vb k_{\text d} $ and the direction of the optical fiber $ \vb k $. For the forward scattering direction, $\theta=0$, which was already calculated in \cite{Kurko2021}, we have 
\begin{multline}
\label{eq:XiGroundMatch}
\xi_0 = \frac{z_\text{R}}{(2\pi)^{3/2}\sigma^2 \sigma_z} \int\dd{V} \frac{1}{q(z)} \exp[-\frac{x^2+y^2}{2}\left( \frac{1}{\sigma^2} + i\frac{k_{\text d}}{q(z)}\right) -\frac{z^2}{2\sigma_z^2} ]  \\
= -i\sqrt{\frac{\pi}{2}} \frac{z_\text{R}}{\sigma_z} \,e^{\frac{\left(z_\text{R}+k_{\text d} \sigma^2 \right)^2}{2\sigma_z^2}}
\mathrm{erfc} \left(\frac{z_\text{R}+ k_{\text d} \sigma^2}{\sqrt{2} \sigma_z}\right) \, \approx \frac{w_0^2}{2\sigma^2+w_0^2} \, ,
\end{multline}
where in last line we again used the approximation that the longitudinal length $ \sigma_z $ of the condensate is much smaller than the Rayleigh range $ z_\text{R} $.
%This is the phase-matched direction $ \vb k_\text{d} = \vb k_{ge} $, which is the only one where significant radiation can be collected from a gas of cold atoms with a distribution function equivalent with that of the  BEC wavefunction (\ref{eq:groundSt}). Not surprisingly, the geometric factors (\ref{eq:XiGround}) in these cases are identical.  However, in \cite{Kurko2021} it was assumed that the amplitude of each atom carries a phase factor $ \exp(i\vb k_\text{d} \vb r) $. \alert{Why was this an assumption? This phase is naturally inscribed by the drive field, isn’t it? Or otherwise I don’t know what you mean…} For ultracold atoms forming a degenerate quantum gas with a macroscopically occupied wavefunction, the phase is fixed across the atomic distribution. The excited state $\ket{e j} \sim  \exp(i\vb k_\text{d} \vb r)$ takes on the incoming photon momentum. Therefore, there is no need for the phase matching to be satisfied to imprint an additional phase pattern into the gas. The degeneracy implies that all atoms forming the condensate radiate in phase, therefore superradiant emission takes place, i.e., there is a dependence $N^2$ on the number of atoms.

So far we have seen that for every $ (\vb k_\text{d}, \vb k_{ge}) $ the radiation collected by the Gaussian optics has two components: one which is independent of the angle $ \theta $, and is proportional to $ N $, and the other which is confined into a small solid angle around $\theta=0$ and is proportional to $ N^2 $. Certainly, this latter dominates the photon emission in the forward direction for sufficiently large atom number. 

\section{Numerical results and discussion}
\label{sec:Discussion}

\begin{figure}
\centering
\begin{subfigure}{.5\textwidth}
  \centering
  \includegraphics[width=.9\linewidth]{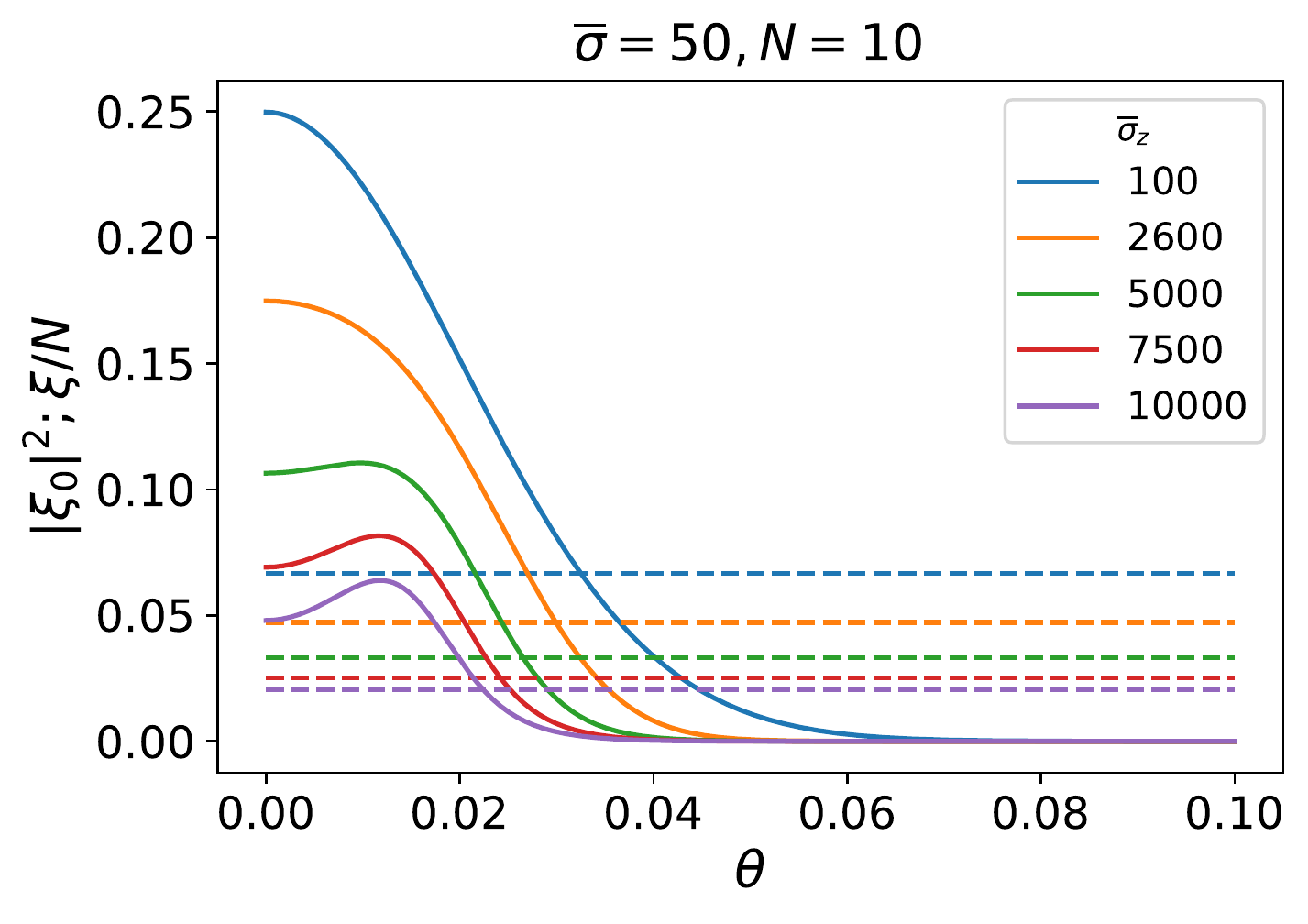}
  \caption{} %The longitudinal size $ \overline{\sigma} = 50  $ is fixed, while the transverse size $ \overline{\sigma}_z $ is varying}
  \label{fig:thetasigz}
\end{subfigure}%
\begin{subfigure}{.5\textwidth}
  \centering
  \includegraphics[width=.9\linewidth]{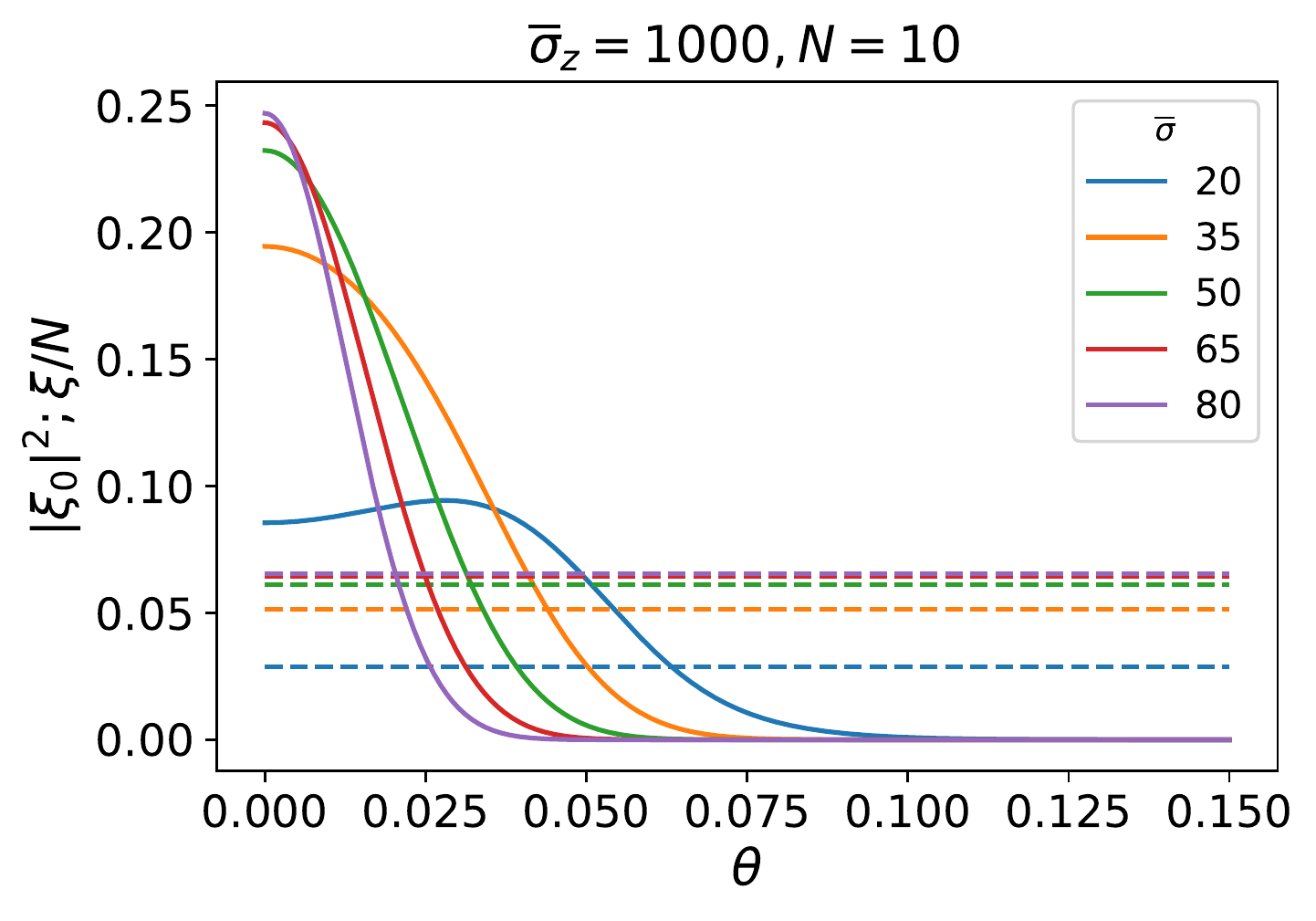}
  \caption{}%The transverse size $ \overline{\sigma}_z = 1000  $ is fixed, while the longitudinal size $ \overline{\sigma} $ is varying}
  \label{fig:thetasig}
\end{subfigure}
\caption{Angular dependence of the geometrical factor $ \abs{\xi_0(\theta)}^2 $ (continuous lines) corresponding to the superradiance and the value  of the geometrical factor $ \xi/N $ (horizontal, dashed lines) for the isotropic radiation for a fixed number of atoms $ N = 10 $, for various values of the transverse size $ \overline{\sigma}_z $ (a) and the longitudinal size $ \overline{\sigma} $ (b)}
\label{fig:theta}
\end{figure}

We consider a cylindrically symmetric harmonic trap and an optical fiber oriented in the $z$ direction, and address the question as to which part of the incoupled intensity, \cref{eq:rate} or \cref{eq:rate0}, gives larger contribution to the collected radiation for a given trap geometry $ \sigma, \sigma_z $ and beam waist $ w_0 $, at different angles $ \theta $. It was already shown in \cite{Kurko2021} that the intensity of the forward scattering is optimized if the beam waist equals $ \sqrt{2}\sigma $. So in the following we are going to fix $ w_0 =\sqrt{2}\sigma $. In the limit of small longitudinal size relative to the Rayleigh range, the geometrical factors \labelcref{eq:Xi} and \labelcref{eq:XiGroundMatch} become $ \xi = 2/3 $ and $ \abs{\xi_0}^2 = 1/4 $, respectively. To compare the two radiation channels, we study the distinct parts of the intensities \labelcref{eq:rate} and \labelcref{eq:rate0}, namely, the geometrical factors $ \abs{\xi_0(\theta)}^2 $ and $ \xi/N $, cf. \cref{fig:theta}, {with dimensionless quantities defined as $ \overline{\sigma} = k_\text{d} \, \sigma $, $ \overline{\sigma}_z = k_\text{d} \, \sigma_z$ and $ \overline{w}_0 = k_\text{d} \, w_0$}.

%%%%%%%%%%%%%%%%%%%%%%%%%%%%
%%%%%%%%%%%%%%%%%%%%%%%%%%

If the BEC has an oblate or a nearly spherical shape, the superradiant part of the geometrical factor takes a maximum value of $ 1/4 $ at the forward direction $ \theta = 0 $, while the isotropic channel results in $ \xi = 2/3 $. In this case $ \overline{\sigma}^2 > \overline{\sigma}_z $, which is equivalent to the setup in which the Rayleigh range is much larger than the longitudinal size $ \sigma_z $. As the spheroid becomes prolate and $ \sigma_z $ becomes comparable to $z_\text{R} $, the geometrical factors of both channels decrease. At a certain point, where the shape of the BEC becomes very elongated, the maximum of the anisotropic radiation is displaced to a nontrivial angle. The reason for this is that the curvature and the Gouy phase term of the Gaussian beam become significant, which can be compensated for in \cref{eq:XiGround} only by a nonzero $ \theta $.

Around $ \theta = 0 $ the dominant contribution comes from superradiant emission from the BEC, but above a particular angle $ \theta^* $, which depends on the trap geometry, the isotropic radiation exceeds the superradiant one. This critical angle increases if the geometrical parameters $ \sigma $ and $ \sigma_z $ decrease.

%%%%%%%%%%%%%%%%%%%%%%%%%%%%
\begin{figure}
\centering
\includegraphics[width=0.6\linewidth]{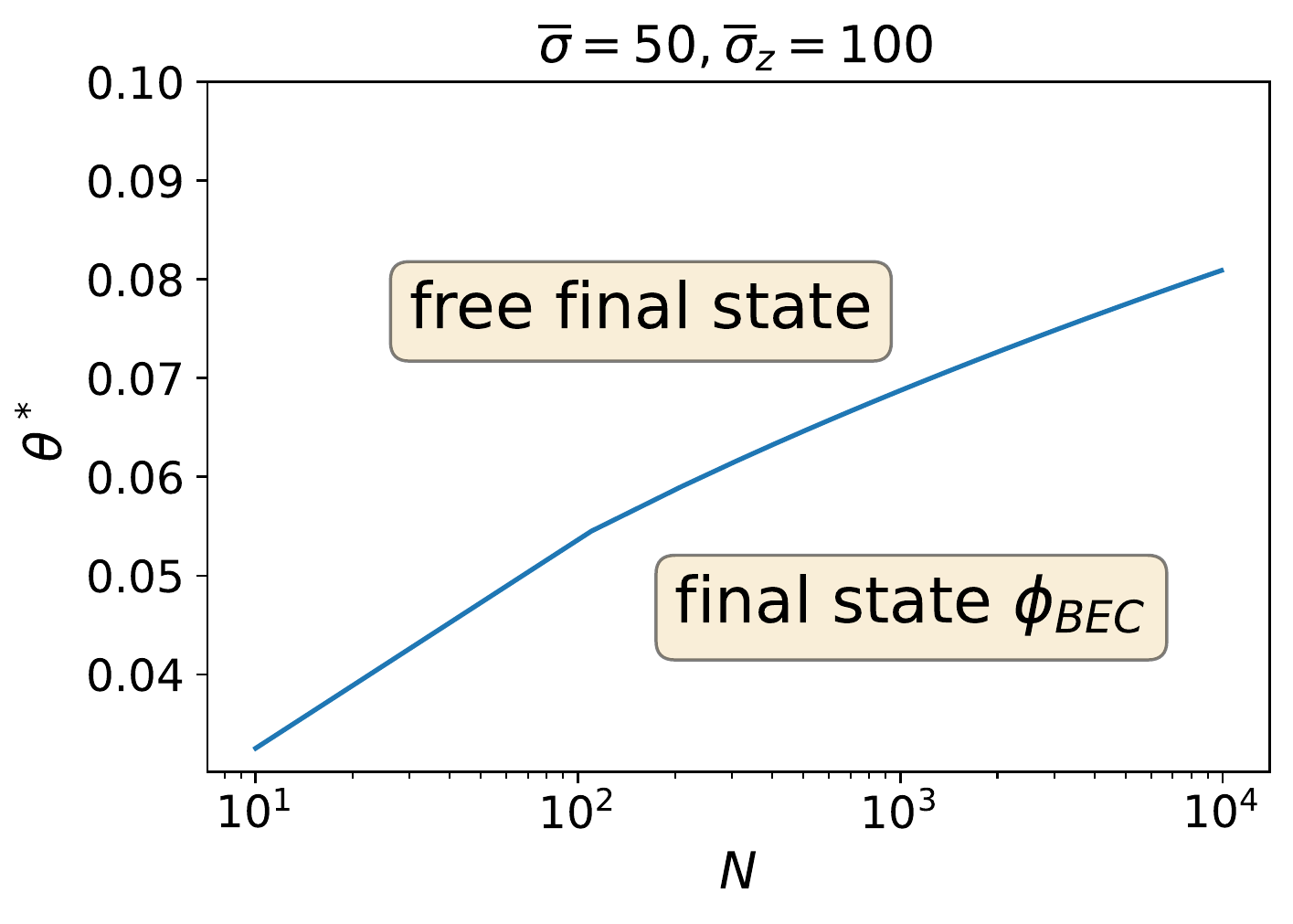}\\
\caption{The dependence of the critical angle $ \theta^* $ (the two channels have the same contribution) on the atom number $N$,
for fixed geometry $ \overline{\sigma} = 50 $, $ \overline{\sigma}_z = 100 $ }
\label{fig:Ndep}
\end{figure}
%%%%%%%%%%%%%%%%%%%%%%%%%%%%%

For small number of atoms, only the isotropic part is significant for nearly all angles $ \theta $, cf. \cref{fig:Ndep}. But as the number of atoms increases, the superradiance becomes important in an ever-widening range around the forward direction, as it is amplified by a factor of $N$ compared to the isotropic radiation.

To summarize, we have considered the photon emission by a Raman process with $\Lambda$-atoms in a BEC. In addition to the internal atomic state, we have taken into account the atomic external, motional degree of freedom in order to describe the momentum transfer in the photon recoil. We performed the analysis in the perturbative, weak excitation limit, in which we were able to separate the spatial dependence of the emitted radiation from its temporal dynamics. Under these approximations, we have shown that there are two channels for the converted radiation: the supperadiant one, which corresponds to the phase-matched photon scattering in the forward direction; and the isotropic one, in which the Bose-Einstein condensate takes away an arbitrary momentum mismatch between the incoming and emitted radiation. The contribution of the latter one is more significant in the side scattering directions, while the width of this region varies with the number of atoms in the condensate and with the dimensions of the harmonic trap.

\section*{Acknowledgement}

The work of A.K., P.D. and A.V. was supported by the National Research, Development and Innovation Office of Hungary (NRDIO) within the Quantum Information National Laboratory. 
DP was supported by the EU QuanERA Project PACE-IN (GSRT Grant No. T11EPA4-00015) and 
by the Alexander von Humboldt Foundation in the framework of the Research Group Linkage Programme.

\bibliography{spp_bec}

\end{document}